\documentclass[12pt]{article}
\usepackage[margin=1in]{geometry}
\usepackage{setspace}
\usepackage{mathtools}
\usepackage{amssymb}
\usepackage{braket}
\usepackage{amsmath}
\usepackage{authblk}
\usepackage[english]{babel}
\usepackage{xcolor}
\usepackage{mathrsfs}

\begin{document}
\begin{center}
\Large{\textbf{Quantum spatial superpositions and the possibility of superluminal signaling}}
\end{center}

\begin{center}
\begin{large}
P. \'Avila, E. Okon, D. Sudarsky and M. Wiedemann\\
\end{large}
\textit{Universidad Nacional Aut\'onoma de M\'exico, Mexico City, Mexico.}\\
\end{center}

A recently proposed gedankenexperiment involving the (gravitational or electromagnetic) interaction between two objects---one placed in a state of quantum superposition of two locations---seems to allow for faster-than-light communication. However, it has been argued that, if the mediating fields are endowed with quantum properties, then the possibility for superluminal signaling is fully avoided. Moreover, in the gravitational case, this conclusion has been used to argue for the view that the gravitational field must be quantized. In this work, we clarify and complement some aspects of the discussion. In particular, by focusing on the way in which entanglement spreads across the components of the system, we offer some insights into the fundamental quantum features behind the impossibility of superluminal signaling and we provide a more general proof of such an impossibility in this and related protocols.

\onehalfspacing

\section{Introduction}

The search for a theoretical framework incorporating general relativity and quantum theory has proven to be one of the most difficult undertakings in physics. A common assumption behind such a pursuit is that gravity itself must have a quantum nature. In fact, the possibility of schemes in which matter fields are treated in quantum terms, but gravity is treated classically, has been argued against on several grounds \cite{eppley1977necessity,page1981indirect}. However, those arguments have been found to be less convincing than intended (see, e.g., \cite{huggett2001quantize,mattingly2006eppley,carlip2008quantum}).
 
What is clear is that the final verdict regarding the fundamental nature of gravity must be informed by experimental evidence arising from situations where both quantum theory and gravitation play a relevant role. The standard expectation is that such situations only emerge in phenomena involving extremely high energies or when curvature values approach the Planck scale (i.e., $ R\sim 1/m_{p}^2$)---both of which are currently well beyond our empirical reach. However, there have been recent proposals that look for a possible quantum behavior of gravity in tabletop experiments, \cite{bose2017spin,Marletto}. In the meantime, there have also been proposals suggesting that useful hints might be acquired by exploring gedankenexperiments involving gravitational fields associated with matter sources in states that require a quantum mechanical treatment, \cite{cecile2011role,zeh2011feynman}.

A concrete instance of this latter approach has been explored in some detail in \cite{mari2016experiments,belenchia2018quantum,danielson}. The gedankenexperiment considered involves two observers: one (Alice) in control of a particle placed in a quantum superposition of two spatial locations and the other (Bob) deciding whether a second particle is allowed to react to its (electromagnetic or gravitational) interaction with the first. The setup is such that an interaction between the particles would seem to prevent the observability of an interference pattern for Alice. And since the decision to allow the particles to interact or not can be taken with spacelike separation between them, the protocol would seem to allow for superluminal signaling.

In \cite{belenchia2018quantum}, however, it is argued that if one attributes quantum properties to the mediating fields, then the possibility for superluminal signaling is fully avoided. In particular, the claim is that taking into account the quantization of radiation of the fields and the existence of their vacuum quantum fluctuations allows for the undesired conclusion to be evaded. Moreover, in the gravitational case, this conclusion is used to argue for the view that the gravitational field must be given a quantum description. More recently, \cite{danielson} attempts a more precise and rigorous evaluation of the entanglement and decoherence effects occurring in the gedankenexperiment. Such an analysis is said to significantly improve upon the rough estimates provided in \cite{belenchia2018quantum} and to show that the conclusions reached in such a work are valid in much more general circumstances.

In this work, we clarify and complement some aspects of the discussions in \cite{belenchia2018quantum,danielson} and we offer a more general proof for the impossibility of superluminal signaling in the scenarios under study. Regarding \cite{belenchia2018quantum}, we first point out that, in order for the signaling protocol to get off the ground, one needs to \emph{presuppose} the quantum nature of the fields. Therefore, the fact that considering the quantum nature of the fields may eliminate the possibility of signaling, cannot be used to argue that gravity must be quantized. Next, by considering a version of the gedankenexperiment in which the second observer controls many particles, instead of one, we show that if one takes the arguments of \cite{belenchia2018quantum} at face value, then it would seem that Bob can in principle send a superluminal signal to Alice. This suggests to us that the explanation offered in \cite{belenchia2018quantum} cannot be taken as fully satisfactory.

Regarding the more precise evaluation presented in \cite{danielson}, we find that, although the path of analysis is certainly a very useful one, the discussion contains a couple of aspects with the potential for generating confusion. Moreover, we point out that the criterion used to argue for the impossibility of superluminal signaling partially obscures the actual element behind such an impossibility. In fact, we will argue that the argument presented in \cite{danielson} can be made more general and transparent by a relatively simple modification. In this regard, by paying attention to the way in which entanglement spreads across tripartite quantum systems in general, and to the spacetime distribution of the different elements of the experimental setting, we show that all observations made by Alice are completely unaffected by any possible interaction (including measurements) between the mediating fields and any systems controlled by Bob---thus explaining away the possibility of superluminal signaling. However, as we shall argue, this does not mean that Bob's equipment has a completely vanishing influence on Alice's particle and the interference patterns she will observe. What is important is that Bob's decisions can have no influence in said patterns, so he is unable to employ his equipment to produce a faster than light communication channel.

Our manuscript is organized as follows. In section \ref{GE}, we describe the gedankenexperiment, as presented in \cite{belenchia2018quantum}, and we quickly review the analyses of such a scenario developed in \cite{belenchia2018quantum,danielson}. Then, in section \ref{EV}, we explore, comment and complement some aspects of such discussions. Next, in section \ref{SP}, by paying attention to the way in which entanglement gets distributed in the gedankenexperiment, we offer some insights into the basic quantum features behind the impossibility of superluminal signaling and present a more general proof for such an impossibility. Finally, in section \ref{Co} we offer our conclusions.

\section{The gedankenexperiment and previous assessments}
\label{GE}

The recent \cite{mari2016experiments} proposed and analyzed a gedankenexperiment in which the (electromagnetic or gravitational) interaction between two objects, one placed in a quantum superposition of two locations, apparently allows for superluminal communication. The experiment was later reanalyzed in \cite{belenchia2018quantum,danielson} (see also \cite{belenchia2019information,wald2020quantum}). Here, we focus on the gedankenexperiment as presented in \cite{belenchia2018quantum} and on the appraisals of the scenario offered in \cite{belenchia2018quantum} and \cite{danielson}.

\subsection{The gedankenexperiment}

The gedankenexperiment considered in \cite{belenchia2018quantum} has two versions, one electromagnetic and one gravitational. Both contain two observers, Alice and Bob, separated by a distance $D$. Alice has control over a particle with spin, charge $q_{A}$ and mass $m_{A}$ and Bob over a particle with charge $q_{B}$ and mass $m_{B}$. In the electromagnetic case, all gravitational effects are ignored; in the gravitational one, the charges are set to zero. We work in units with $\hbar = c = 1$.

The experiment starts by assuming that, in the distant past, Alice's particle was sent through a Stern-Gerlach apparatus, leaving its state in the superposition $\frac{1}{\sqrt{2}} (\ket{L}_{A} \ket{\downarrow}_{A}+\ket{R}_{A} \ket{\uparrow}_{A})$ with distance $d$ between $\ket{L}_{A}$ and $\ket{R}_{A}$. This step is assumed adiabatic, with negligible radiation emitted. Bob's particle, on the other hand, is initially assumed to be held on a strong trap, so its interaction with Alice's particle is negligible. The experiment then proceeds as follows. At time $t=0$, Bob decides whether to release his particle from the trap or leave it there; we call $T_B$ the time at which Bob completes his experiment. Also at $t=0$, Alice starts an interference experiment with her particle, which ends at time $T_A$ (see Figure 1). 
\begin{figure}[!ht]
\centerline{\includegraphics[scale=0.325]{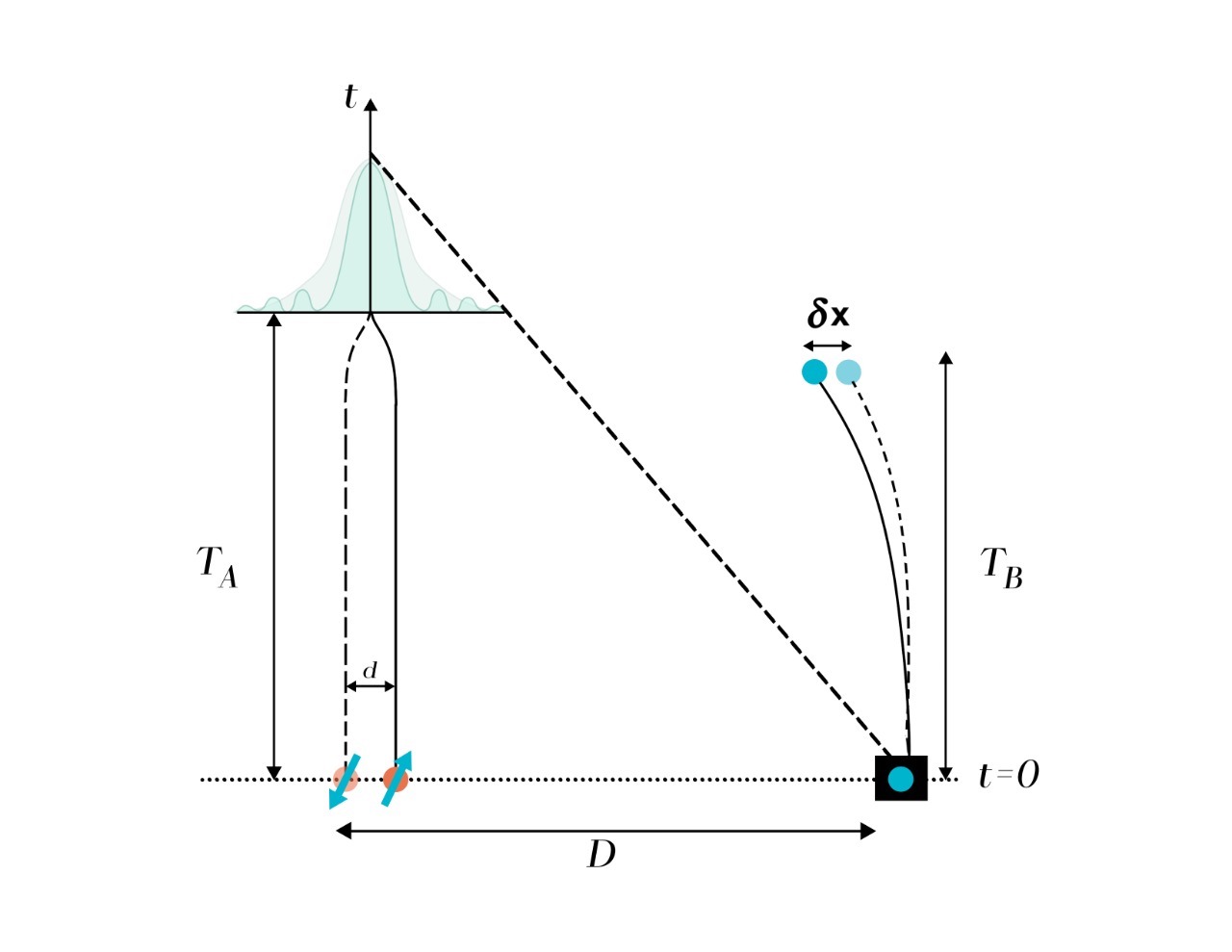}}
\caption{Spacetime diagram of the gedankenexperiment.} 
\end{figure}

The experiment is then analyzed as follows. If Bob decides to release his particle, it would get entangled with the fields produced by the components $\ket{R}_{A}$ and $\ket{L}_{A}$ of Alice's particle, which would put Bob's particle in a spatial superposition with separation $\delta x$. If $\delta x$ is large enough, the states of that superposition would be nearly orthogonal, so Bob's particle would be nearly maximally entangled with Alice's, which would prevent her from observing interference. If, on the other hand, Bob does not release his particle, nothing would prevent Alice from observing interference. It is then concluded that Bob's decision to release or not the particle determines whether Alice observes interference or not. Therefore, if $T_{A}, T_{B} < D$, in which case the experiments of Alice and Bob would be spacelike separated, Alice and Bob would seem to have access to a superluminal channel.\footnote{In \cite{belenchia2018quantum}, this result is presented as a tension between complementarity and causality: if complementarity holds, superluminal signaling would be possible.} 

\subsection{A proposed solution via the quantum nature of the fields}

According to \cite{belenchia2018quantum}, the apparent possibility of superluminal signaling described above is fully avoided if the quantization of radiation of the mediating fields and the existence of their vacuum quantum fluctuations are taken into account. Regarding the quantization of radiation, it is noted that, for Alice to be able to coherently recombine her particle, she must do the experiment in a time $T_A$ long enough to avoid the emission of even one photon, in the electromagnetic case, or one graviton, in the gravitational one. Estimates of the amount of radiation emitted lead to the conditions
\begin{equation}
\label{AC}
\mathcal{D}_A<T_A
\quad \textup{(electromagnetic case)}\qquad\quad \mathcal{Q}_A<T^2_A
 \quad \textup{(gravitational case)}.
\end{equation}
with $\mathcal{D}_A$ Alice's ``effective dipole moment'' and $\mathcal{Q}_A$ its ``effective quadrupole moment''.

In addition, it is argued that the inevitable vacuum fluctuations of the fields impose a necessary delocalization of particles interacting with them---charge-radius in the electromagnetic case and Planck length in the gravitational one. Given this, it is argued that, in order for Bob to be able to destroy Alice's coherence, he would need for the displacement of his particle to be larger than its delocalization. This is estimated to occur when
\begin{equation}
\label{BC}
\frac{\mathcal{D}_A}{D^3}T_B^2>1
\quad \textup{(electromagnetic case)}\qquad\quad \frac{\mathcal{Q}_A}{D^4}T_B^2>1
 \quad \textup{(gravitational case)}.
\end{equation}

Now, for Bob to be able to influence Alice's interference experiment superluminally, we need $T_A < D$ and $T_B < D$. However, if (\ref{AC}) holds, then $\mathcal{D}_A<D$ and $\mathcal{Q}_A<D^2$ so, from (\ref{BC}), that means that Bob cannot disrupt Alice's interference in time for signaling to occur. With all this, it is concluded that, by postulating the quantization of radiation and the existence of quantum vacuum fluctuations of the gravitational field, the worry of superluminal signaling is completely avoided. Moreover, in the gravitational case, such a conclusion is taken as providing support for the idea that the gravitational field must be described in quantum terms.

\subsection{A more general analysis of the decoherence effects}

In \cite{danielson}, a more rigorous analysis of the gedankenexperiment is attempted. The new analysis is taken as significantly improving on the rough estimates in \cite{belenchia2018quantum} and as showing that the conclusions of such a work are valid in much more general scenarios.

After some groundwork offering a precise description of the entanglement and decoherence effects involved, \cite{danielson} considers a Cauchy surface, $\Sigma_1$, containing the recombination event, but to the past of all of Bob's experiment and another Cauchy surface, $\Sigma_2$, for which Alice has not started the recombination, but Bob has completed his measurement. After this, they point out that it is possible for more decoherence to occur as Alice recombines her particle. However, it is argued that, since Bob has completed his experiment, he has stopped interacting after $\Sigma_2$, so it is impossible for the decoherence of Alice's particle to be less than the decoherence caused by Bob. That is, denoting by $\ket{B_1}$, $\ket{B_2}$ the states of Bob's apparatus after completing his experiment and $\ket{\Psi_1}$, $\ket{\Psi_2}$ formally corresponding to the fields generated by Alice's particle, it is argued that if 
\begin{equation}
\label{Par}
\lvert \braket{B_1|B_2} \rvert < \lvert \braket{\Psi_1|\Psi_2}_{\Sigma_1} \rvert ,
\end{equation}
could be the case, then a paradox would arise. Moreover, it is argued that, if Eq. (\ref{Par}) would hold, then Bob's measurement either would violate causality or it would violate complementarity. In other words, it is argued that Eq. (\ref{Par}) is a precise statement of the potential paradox posed by the gedankenexperiment.

However, according to \cite{danielson}, it is easy to see that such a situation simply cannot arise. To show it, \cite{danielson} starts by considering the state of the whole system at $\Sigma_1$ and examines its evolution to $\Sigma_3$, a spacelike hypersurface that lies to the future of, both, Alice's recombination event and Bob's experiment. Then, by pointing out that such an evolution is unitary, so norms of states are preserved, they argue that
\begin{eqnarray}
\label{AB}
\braket{\Psi'_1|\Psi'_2}_{\Sigma_3} \braket{B_1|B_2} & = & \braket{\Psi_1|\Psi_2}_{\Sigma_1} \braket{B_0|B_0} \nonumber \\
& = & \braket{\Psi_1|\Psi_2}_{\Sigma_1} ,
\end{eqnarray}
(with $\ket{\Psi'_1}_{\Sigma_3}$ and $\ket{\Psi'_2}_{\Sigma_3}$ the radiation states after interaction with Bob) from which it follows that 
\begin{equation}
\lvert \braket{B_1|B_2} \rvert \ge \lvert \braket{\Psi_1|\Psi_2}_{\Sigma_1} \rvert .
\end{equation}
That is, it is concluded that the inequality of Eq. (\ref{Par}) can never be satisfied, thus avoiding the possibility of using this setup as a faster-than-light signaling protocol.

\section{An evaluation of previous assessments}
\label{EV}

In this section, we scrutinize some aspects of the discussions in \cite{belenchia2018quantum,danielson}. First, we examine the claim in \cite{belenchia2018quantum} that the fact that ascribing quantum properties to the gravitational field helps avoid superluminal signaling, could be taken as lending support to the idea that the gravitational field must be quantized. Next, we consider in detail a scenario in which Bob controls many traps to show that, if the explanations offered in \cite{belenchia2018quantum,danielson} where all there is to the issue, then it would not be completely clear, from an heuristic standpoint, why signaling could not occur. Finally, we quickly comment on another potentially problematic aspect of the analysis in \cite{belenchia2018quantum}, having to do with the use of \emph{center of mass} wave functions, and we close with the discussion of some aspects of the more general analysis in \cite{danielson} which, we argue, could lead to misunderstandings.

\subsection{On the quantization of the gravitational field}
\label{qnf}

We start by scrutinizing the claim in \cite{belenchia2018quantum} that endowing the gravitational field with quantum properties avoids superluminal signaling lends support to the idea that the gravitational field must be given a quantum description. We note, however, that the signaling protocol of the gedankenexperiment \emph{presupposes} that the mediating fields have a quantum nature. Therefore, the use of the protocol to argue that gravity must be treated quantum mechanically must be regarded as circular. 

It should be emphasized that we are not advocating for the position that the gravitational field must \emph{not} be given a quantum treatment. As far as we know (and, in fact, this is something we consider very likely), it might be the case that the gravitational field should be treated in quantum terms. The point we are objecting is the use of this gedankenexperiment, together with the natural expectation that no superluminal signaling is allowed by nature, as an argument in favor of such a view.

In detail, the issue is that the description of the signaling protocol takes as a given that the field in question (gravitational in this case) has quantum properties. To see this, we note that it is because one assumes that the field generated by the superposition of Alice's particle is also described by a corresponding quantum superposition of states---that is, that the gravitational field gets entangled with Alice's particle---that Bob's particle also ends up entangled with Alice's particle. Otherwise, Bob's particle would not get entangled with Alice's particle and would be unable to destroy Alice's interference pattern. It is important to point out that this should not be read as implying that the quantum nature of the field is a necessary condition for generating entanglement. For example, in \cite{doner2022gravitational} it is argued that a pilot-wave inspired model, in which gravity is given a classical treatment, is able to produce entanglement.

We should note the possibility of some sort of semiclassical description, in which matter fields are treated quantum mechanically, the treatment of gravity remains classical, and where the geometry associated with the delocalized quantum superposition of Alice's particle corresponds to the geometry associated with a sort of \emph{average} of the two terms of the superposition. In that case, Bob's particle would simply respond to that single ``average field'' and not get entangled with Alice's particle. That is, if the Einstein curvature tensor for the spacetime metric is being sourced by the expectation value of the energy-momentum tensor, then no entanglement of the sort assumed in the very conceptualization of the communication protocol will ever be generated.

One could argue that semiclassical frameworks of this sort are already ruled out by works such as \cite{eppley1977necessity,page1981indirect}. However, as we mentioned above, such arguments are not as conclusive as intended.\footnote{Even though existing arguments against semiclassical gravity are not conclusive, they do set constraints on semiclassical frameworks (see also \cite{Tim} for a general assessment of the issue).} In any case, the argument in \cite{belenchia2018quantum} is supposed to be one against semiclassical gravity, so it cannot depend on assuming that semiclassical gravity is not viable.\footnote{Our position in this regard is that it might very well be the case that spacetime only allows for a classical description. One extreme possibility is that spacetime is classical at the fundamental level. But there is another reasonable option, which is that, even though the fundamental description is fully quantum, spacetime is emergent and, by the time that it emerges, only a classical description is available. Claims to the contrary are problematic, not only from a mere academic perspective, but could also conceal the potential utility of semi-classical frameworks in regimes where quantum aspects of matter play essential roles.}

In sum, given that the protocol presupposes the fields to be quantum, the fact that their quantum nature may be used to avoid signaling should, at best, be read as a self-consistency proof and cannot be used to argue for the need of quantization of the field in the first place. At most this can be said to support the view that, if one considers that somehow a spatial superposition of quantum sources is associated with a quantum superposition of states of the gravitational field, then one must include consideration of the quantum uncertainties in the state of such fields in order to avoid being mislead into erroneous conclusions.

\subsection{Multiple Traps}

In this section we first argue that, if one takes at face value the arguments made in \cite{belenchia2018quantum}, and apply them to a system in which Bob controls $N$ traps, instead of only one, then one would conclude that Alice's system could suffer measurable decoherence---opening the door for superluminal signaling. We then address a possible response to our argument offered in \cite{belenchia2018quantum}, involving vacuum fluctuations, and explain why we find this solution to be unsatisfactory. Finally, we explain why the solution put forth in \cite{belenchia2018quantum} cannot be the whole story and take a few first steps towards a more general resolution.

\subsubsection{The $N$-traps scenario}
According to \cite{belenchia2018quantum}, for Bob to be able to disrupt Alice's interference experiment, the displacement $\delta x$ of his particle must be larger than the width of its wave function. However, this way of expressing the issue could lead to misunderstandings. The point is that interference is not an all or nothing affair and all that is required for Bob to be able to send a signal to Alice is for Bob's experiment to noticeably affect Alice's interference experiment. That is, as long as the interference pattern observed by Alice without Bob's experiment is different from that obtained when Bob performs his experiment, a signal could be sent. Therefore, even if the displacement of Bob's particle is not larger than the width of its wave function, a signal could be sent.

The issue, in other words, is that the analysis in \cite{belenchia2018quantum} could be interpreted as implying that Bob can cause some decoherence on Alice's particle, only not enough to send a signal. That problem with this is that it opens up the possibility of Bob sending a signal by employing many traps to enhance the minute decoherence he is able to cause with one particle. To see this, consider a scenario in which Bob, instead of controlling one particle, controls $N$ of them. As we will see, in that case it would seem that Bob can cause decoherence on Alice's experiment at time $T_B<D$, even if $\mathcal{D}_A<T_A<D$. That is, that he would be able to send a superluminal bit of information to Alice---even if the quantization of the field and its fluctuations are fully taken into account.

According to the analysis in \cite{belenchia2018quantum}, the vacuum fluctuations of the fields induce a limit on the localization of a particle in them. Such limits are argued to be given by $\sigma=q/m$, in the electromagnetic case, and by $\sigma=l_P$ in the gravitational one. As a result of this, it is argued that, in order for Bob to be able to decohere Alice's particle, the separation in his superposition, $\delta x$, must be larger than $\sigma$. Finally, it is claimed that, if Alice recombines her particle without emitting radiation, Bob will not be able to obtain a $\delta x$ large enough, in time to disrupt Alice's recombination. 

Suppose, then, that Alice does recombine her particle without emitting radiation. In that case, when Bob releases his particle, he will obtain a $\delta x$ smaller than $\sigma$, so the inner product of the states $\ket{L}_B$ and $\ket{R}_B$ of Bob's superposition will satisfy
\begin{equation}
|\braket{L|R}_B|=1-\epsilon
\end{equation}
for some $\epsilon \ll 1$.

Suppose, now, that Bob has not one, but $N$ particles. If he decides to release them, and assuming no interaction between the $N$ particles, their state becomes
\begin{equation}
\label{ST}
\ket{\Psi_N}=\frac{1}{\sqrt{2}}(\ket{L}_A \ket{L_1}_{B_1} \ket{L_2}_{B_2}... \ket{L_N}_{B_N}+\ket{R}_A \ket{R_1}_{B_1} \ket{R_2}_{B_2}... \ket{R_N}_{B_N}) .
\end{equation}
Considering the same conditions for the $N$ particles as we had for the single particle above, the inner product of the right and left components of Bob's particles then becomes
\begin{equation}
\label{N- Ent}
\Pi_i |\braket{L_i|R_i}_{B_i}|=(1 - \epsilon)^N \approx 1 - N \epsilon.
\end{equation}
It seems, then, that if $N$ is large enough, this inner product approaches zero, so Bob's particles become able to cause decoherence of Alice's particle. And this is so, even maintaining the experiments of Alice and Bob at a spacelike distance and with Alice avoiding her particle to radiate. That is, the analysis in \cite{belenchia2018quantum} seems to suggest that, for a large enough $N$, a superluminal communication channel between Alice and Bob would still be possible.

\subsubsection{The role of vacuum fluctuations}
In \cite{danielson}, the possibility of using $N$ traps is explicitly considered, but it is dismissed on the grounds that ``vacuum fluctuations over spacelike separated regions are correlated, so it is not obvious that the $N$ experiments can be treated as independent''. The idea being that, due to correlations between vacuum fluctuations over spacelike separated regions, it would not be correct to assume the $N$ particles to be independent, so it would be incorrect to take their state to be given by either of the terms in the right hand side of Eq. (\ref{ST}).\footnote{The point (stressed to us in a private communication with the authors of \cite{danielson}) is that, as a result of those ``vacuum fluctuations", the $N$ particles would become entangled in complicated ways, so as to completely invalidate the estimate for the increased decoherence reflected in (\ref {N- Ent}).} We do not believe, though, this to be the reason why an $N$-particle protocol would not work.

To begin with, while it is true that vacuum fluctuations over spacelike separated regions are correlated, and that this would cause initially independent systems to become entangled as a result of their interaction with the vacuum, such an effect is believed to be significant only for systems localized to a better degree than their charge-radius. However, the $N$-particle protocol described above, does not seem to require any of the systems to be that well localized. Moreover, the entangling effect of the vacuum fluctuations rapidly decays with distance (the two-point function for the electric field in the vacuum decays as the fourth power of the inverse distance) and there seem to be, in fact, possible arrangements that maintain all of Bob's particles as far from each other as one likes, making it reasonable for their entanglement to be neglected. The point is that, for any given desired minimum distance between Bob's particles, $r$, there are arrangements for Bob's particles with no pair closer than $r$. To see this, we note that the decoherence caused by a set of particles arranged over a sphere, with a given density of particles per unit area, can be made to be independent of the radius of the sphere. It is clear, then, that by using an appropriate density of particles per unit area and an appropriate number of spheres, the decoherence caused by Bob can be made as large as one would like, with no pair of particles closer than $r$ (see the appendix \ref{Ap} for explicit calculations).

\subsubsection{The role of entanglement}
We just saw that there are arrangements for Bob's particles that make it reasonable to assume that they are independent. It seems to us, though, that not much would change, even if Bob's particles were allowed get entangled---through vacuum fluctuations or some other mechanism. To see this, suppose that Bob has $N \gg 1$ entangled particles in the state
\begin{equation}
\ket{\Psi}= \sum_{i} a_{i} \ket{\psi^{i}},
\end{equation}
with the $\ket{\psi^{i}}$ separable states of the $N$ particles. As the particles interact with the field produced by Alice's particle, their state evolves to 
\begin{equation}
\ket{\Psi}= \frac{1}{\sqrt{2}} \left( \ket{\Psi_{L}}+ \ket{\Psi_{R}} \right),
\end{equation}
with $\ket{\Psi_{L}}= \sum_{i} a_{i} \ket{\psi^{i}_{L}}$ and $\ket{\Psi_{R}}= \sum_{i} a_{i} \ket{\psi^{i}_{R}}$. That is, the $\ket{\psi^{i}_{L}} (\ket{\psi^{i}_{R}})$ are the result of the interaction between the $i$-th separable state of the $N$ particles and the left (right) component of the field produced by Alice's particle.

Consider now $\braket{\Psi_{L}|\Psi_{R}}=\sum_{ij} a_{i} a_{j} \braket{\psi_{L}^{i}|\psi^{j}_{R}}.$ Given what we saw above regarding the case of $N$ non-entangled particles, we have that $\braket{\psi^{i}_{L}|\psi^{i}_{R}} \approx (1-\epsilon)^{N}$. What about the crossed terms $\braket{\Psi_{L}^i|\Psi_{R}^j}$? We would like to argue that it is reasonable to assume that $\braket{\psi^{i}_{L}|\psi^{j}_{R}} \approx 0$. The point is that, it would be an amazing coincidence for the state $\ket{\psi^i}$, influenced by the Alice's particle to the left, to have a significant overlap with $\ket{\psi^j}$, influenced by the Alice's particle to the right. Remember that the separable states in question are $N$-particle states, with $N \gg 1$, so it is enough for one of the particles not to have overlap for the whole inner product to vanish. 

Putting everything together,
\begin{equation}
\braket{\Psi_{L}|\Psi_{R}} \approx \sum_{i} |a_{i}|^{2} \braket{\psi^{i}_{L}|\psi^{i}_{R}} \approx (1-\epsilon)^{N} \sum_{i} |a_{i}|^{2}=(1-\epsilon)^{N}.
\end{equation}

As a simple illustration, let us consider a two-particle system constrained to a plane, in the entangled state 
\begin{equation}
\ket {\Psi} = \frac{1}{\sqrt{2}} \left[ \psi ( \vec x_1- a\hat x ) \psi( \vec x_2 +a\hat x ) + \psi ( \vec x_1- a\hat y ) \psi( \vec x_2 +a\hat y) \right] ,
\end{equation}
with $ \hat x$ the unit vector in the  $x$-direction and $ \psi(\vec x) $ a Gaussian wave packet of width $ \delta \ll a$ centered at the origin. Consider now the overlap of such a state with $\ket {\Psi_\varepsilon}$, given by a translation of $\ket {\Psi}$ by a small quantity $\varepsilon \ll \delta$ along $x$. As expected,
\begin{equation}
\braket{\Psi|\Psi_\varepsilon }=e^{-\frac{2a^2+\varepsilon^{2}}{4\delta^{2}}} +e^{-\frac{\varepsilon^{2}}{4\delta^{2}}} \approx e^{-\frac{\varepsilon^2}{4\delta^2}} \approx (1-\epsilon)^2
\end{equation}
for $\epsilon = \varepsilon^2/8 \delta^2$. We conclude that entanglement need not diminish the decoherence caused on Alice by $N$ particles.

\subsubsection{Towards a more general explanation}
In sum, the analysis of the experiment offered in \cite{belenchia2018quantum} seems to suggest that Bob is able to cause some decoherence on Alice, only not enough of it for a signal to be sent. This seems to open the door for an alternative protocol in which, controlling $N$ traps, Bob would be able to send a superluminal signal. However, according to \cite{danielson}, such an $N$-trap protocol would not work because correlations between vacuum fluctuations would inevitably entangle the different traps. Above we have given what we take to be convincing evidence that vacuum fluctuations are not the reason why $N$-trap protocol would fail. In fact, we are doubtful that vacuum fluctuations are the reason behind the impossibility of signaling. Why is it, then, that all of these protocols do not allow for a superluminal signal to be sent?

A clue could arise by paying attention to the role played by the traps. The issue is that the trap that maintains Bob's particle at its place, in the presence of the field associated with Alice's particle, must be in a state as to compensate the different forces exerted by the field---and thus must get entangled with such a field. To make things concrete, and focusing on the electromagnetic case, let us think of the trap as a Faraday box. Such box effectively shields a charged particle located inside from the electric field associated with external charges. In order for the shielding to work, the walls of the box need to be very good conductors, so the electric charges of the material they are made of would distribute themselves in such a manner as to cancel the field of Alice's particle. The point is that, for each of the two superimposed paths that Alice's particle is following, the resulting configuration of charges would be rather different. Therefore, even before Bob releases his particle, Alice's particle would be entangled with the trap, potentially causing decoherence. It seems, then, that Bob's decision to release or not the particle (or particles) would only exchange the existing entanglement between Alice's particle (and its field) and the trap, to a corresponding entanglement with Bob's particle (or particles). Actually, even considering a situation with no trap, the analysis in \cite{belenchia2018quantum} seems not to be paying attention to the potential decoherence that the field itself could cause on Alice's particle. 

In fact, according to \cite{belenchia2018quantum}, it will typically be the case that the states of the field associated with the left and right paths of Alice's particle would be almost orthogonal. Therefore, in this sense, Alice's particle will have decohered, even before Bob could make a decision. However, \cite{belenchia2018quantum} argues that this would be a case of what \cite{unruh2000false} calls ``false decoherence'': if Alice recombines her particle adiabatically, then the fields would ``follow'' the particle and would allow for a complete recombination. However, this cannot be the whole story. As is shown below, that fact is that the field could cause some decoherence on Alice's particle and, moreover, that interaction between the field and Bob's particle is fully unable to cause on Alice's particle more decoherence than what the field was causing initially. That is, it is not that Bob can cause some decoherence, only not enough of it to send a signal. Instead, Bob, through his actions, cannot cause any decoherence at all. We see, then, that it is not the vacuum fluctuation that disallow for the $N$-trap protocol to work. Instead, given that even a single particle cannot cause any additional decoherence, having $N$ of them does not improve the situation in any way.

\subsection{The center of mass wave function}

Another potentially problematic aspect of the analysis in \cite{belenchia2018quantum} we quickly comment on is the fact that the estimate of the decoherence that Bob can cause on Alice is estimated by considering the overlap of the wave functions of the \emph{center of mass} of Bob's system, which is claimed to possibly be a macroscopic object. The issue is that it is easy to see that different configurations of a collection of particles could share a very similar center of mass wave function, but still be essentially orthogonal. Therefore, the fact that the center of mass wave function causes insignificant decoherence says very little regarding the decoherence caused by the whole object. As a simple example, consider again a two-particle system on a plane, in either of the following states
\begin{equation}
\psi ( \vec x_1- a\hat x ) \psi( \vec x_2 +a\hat x ) \qquad \text{or} \qquad \psi ( \vec x_1- a\hat y ) \psi( \vec x_2 +a\hat y ) ,
\end{equation}
again, with $ \hat x$ the unit vector in the  $x$-direction and $ \psi(\vec x) $ a Gaussian wave packet of width $ \delta \ll a$ centered at the origin. For both states, the center of mass has an expectation value $X_{cm} = 0$ and uncertainty $\Delta X_{cm} \sim \sqrt{ 2}\delta$. Therefore, one could expect that their overlap would be rather large. However, the inner product of these two states is given by
\begin{equation*}
 \braket{ \Psi^{b}|\Psi^{a}}= e^{-\frac{a^{2}}{4\delta^{2}}},
\end{equation*}
which almost vanishes. Thus, one ought to worry whether the back-of-the-envelope analysis of this question, as presented in \cite{belenchia2018quantum}, could be considered reliable.

\subsection{Possible misunderstanding with the more general analysis}

The more general analysis in \cite{danielson} undoubtedly represents an important addition to the one in \cite{belenchia2018quantum} and contains valuable elements that have played a key role in our own considerations. Still, we find that the discussion has certain features that could lead to misunderstandings. 

We start by recalling that, according to \cite{danielson}, the inequality
\begin{equation}
\label{Par2}
\lvert \braket{B_1|B_2} \rvert < \lvert \braket{\Psi_1|\Psi_2}_{\Sigma_1} \rvert 
\end{equation}
represents a precise statement of the potential paradox posed by the gedankenexperiment. There are, however, a couple of issues about this claim that could prove confusing. The first one is that the whole argument in favor this inequality is carried out with to particular spacelike hypersurfaces in mind: $\Sigma_1$, containing Alice's recombination event, but to the past of all of Bob's experiment and $\Sigma_2$, for which Alice has not started the recombination, but Bob has completed his measurement. The problem is that the inequality simply is not generically true for those hypersurfaces. The issue is simply remedied by considering the decoherence of Bob, not on $\Sigma_2$ but on $\Sigma_3$, containing the recombination event and to the future of Bob's experiment. This, in fact, is what is actually done in \cite{danielson} to prove the inequality (see Eq. (4.5)): if one considers hypersurfaces $\Sigma_1$ and $\Sigma_3$, both passing through the recombination event, then the evolution of Alice's particle is trivial and Eq. (\ref{Par2}) follows.

The second, more important issue, is that we are doubtful of the claim that it is the inequality which captures the possibility or impossibility of signaling. The point is that the decoherence caused by Bob being smaller or larger than that caused by Alice does not seem to have anything to do with Bob's ability to send a signal. In other words, even if the decoherence caused by Bob is smaller than that caused by Alice, Bob could, in principle, use the effect to send a superluminal signal. As we saw above, interference is not an all-or-nothing issue and all that is needed for Bob to signal Alice is for Bob's experiment to perceptibly alter the results of Alice's interference experiment. That is, even if Alice causes some decoherence, and even if the decoherence caused by Bob is smaller than that caused by Alice, as long as the interference pattern obtained by Alice, in the absence of Bob's experiment, is noticeably different from the pattern obtained when Bob performs his experiment, Bob would be able to communicate with Alice. 

We conclude that Eq. (\ref{Par2}) is not the most straightforward way of capturing a potential for signaling. We do recognize, though, that the derivation of the inequality contains a key element in the proof for the impossibility of signaling in the scenario under study, namely,
\begin{eqnarray}
\braket{\Psi'_1|\Psi'_2}_{\Sigma_3} \braket{B_1|B_2} & = & \braket{\Psi_1|\Psi_2}_{\Sigma_1} \braket{B_0|B_0} \nonumber \\
& = & \braket{\Psi_1|\Psi_2}_{\Sigma_1} ,
\end{eqnarray}
from which it is clear that the decoherence caused by Bob and the field on $\Sigma_3$ is exactly equal to that caused by the field alone on $\Sigma_1$. In other words, Bob is simply unable to alter the decoherence observed by Alice. We note, however, that this analysis assumes a purely unitary evolution at all times, which is an important limitation, as Bob could perform measurements on its particle, e.g., to acquire which-way information. Moreover, the analysis presupposes that Alice's particle is initially unentangled with Bob's particle and its trap, which, as we argue below, might not be a good representation of the physical situation in the lab.

In the next section, we provide a proof of the impossibility of superluminal signaling that relaxes the purely unitary evolution assumption, as well as the restriction concerning initial entanglement. Moreover, given that the proof is fairly abstract, we feel it convenient to offer some insights into the fundamental mechanism behind the proof.

\section{A more general proof and some insights}
\label{SP}

In this section, we offer a more general proof of the impossibility of superluminal signaling between Bob and Alice in the gedankenexperiment and related scenarios. We show that, in contrast with what is argued in \cite{belenchia2018quantum,danielson}, it is not vacuum fluctuations and quantization of radiation which play essential roles, but a more basic feature of the way in which entanglement spreads across the components of tripartite quantum systems.

We start by recalling that, according to \cite{belenchia2018quantum}, when Alice recombines her particle slowly, the fields corresponding to the two paths of her particle undergo so-called ``false decoherence''. The idea is that, when Alice's particle is in a superposition of two locations, the fields corresponding to the two components of the superposition are nearly orthogonal. However, when Alice's particle is adiabatically recombined, then the fields ``follow'' the particle, allowing for a perfect recombination. The problem is that the analysis cited to support such a claim, \cite{unruh2000false}, is performed in a setting with either a spin or an harmonic oscillator coupled to a \emph{massive} scalar field. Moreover, the analysis is carried out in the regime in which the time scale of the behavior of the system is larger that the inverse of the mass of the field. It seems, then, that if one wants to apply the conclusions of such an analysis to the gedankenexperiment in question, in which the fields are \emph{massless}, then one would have to consider the case in which the recombination of Alice's particle takes an infinite time to complete. If, on the other hand, the recombination takes a finite time---as demanded by the gedankenexperiment---then the conclusion in \cite{unruh2000false} simply does not carry over.

In fact, it seems clear to us that the field would not simply ``follow'' Alice's particle as she recombines it. After all, in regions sufficiently far away from the recombination event, the state of the field would not have had time to change as a result of the recombination, precisely because such information is expected to travel causally. Any failure of the previous expectation would seem to offer, by itself, a path for superluminal signaling. It is true that, when Alice recombines her particle, a sort of ``partial false decoherence'' occurs. That is, there is a part of the field, e.g., the ``Coulomb part'' defined in \cite{danielson}, which, by definition, does follow Alice's particle. The rest of the field, though, would not undergo any sort of ``false decoherence''. It seems clear, then, that the only way in which Alice would ever be able to fully recombine her particle, is when these "non-Coulomb" parts of the fields, corresponding to the two components of the superposition of her particle, are \emph{never} nearly orthogonal. Otherwise, the decoherence they produce would still be present on any spacial hypersurface containing the recombination of Alice's particle and it would always prevent interference.

Is the above realization enough to explain away the possibility of superluminal signaling? It would seem that it is not, because, even if Alice performs the experiment in such a way that the associated fields only produce negligible decoherence, interaction between the fields and surrounding matter would seem to be able to amplify such decoherence. If so, presence or absence of matter surrounding Alice's experiment would change her observations, opening the door for superluminal signaling. However, as we show below, \emph{any interaction between the fields and any other objects is simply unable to increase the amount of decoherence caused on Alice's particle}. Therefore, if the fields produce negligible decoherence, which, as we saw, is something required for Alice to obtain an interference pattern, interaction between the fields and anything else would not alter in any way the pattern observed by Alice, shutting the possibility of any sort of superluminal signaling. 

To see that interaction between the fields and other elements cannot increase the amount of decoherence caused on Alice's particle, we employ an adaptation of the so-called no-signaling theorem, \cite{NST}. In more detail, together with \cite{belenchia2018quantum}, we treat the particles with non-relativistic quantum mechanics and the mediating fields as a relativistic quantum fields. Moreover, as in \cite{danielson}, we assume that, at least formally, there is a Hilbert space for the combined system. Next, we consider spacelike hypersurfaces $\Sigma_1$, containing the recombination of Alice's particle, but to the past of ``Bob's experiment'', and $\Sigma_3$, also containing the recombination of Alice's particle, but to the future of ``Bob's experiment''; we use the quotation marks to denote the fact that we are generalizing the scenario by considering the interaction between the fields and, not only Bob's particles, but any objects interacting with the fields.

Now, regarding the evolution from $\Sigma_1$ to $\Sigma_3$, we note that, since both hypersurfaces go through the recombination event, the evolution of Alice's particle would be the identity. As for the evolution of the mediating field ($F$) and surrounding objects ($B$), we allow for their interaction to be as general as possible, so we model it as a general quantum operation acting on the $F-B$ system---which include, among other things, purely unitary evolution and non-selective measurements. That is, in contrast with previous analysis, we are contemplating any operation that Bob could locally perform on his system, including measurements on any observables, not limiting our consideration to unitary evolution. General quantum operations can be written as 
\begin{equation}
\mathcal{O}_{FB}(\rho_{FB}) = \sum_i K_i^\dag \rho_{FB} K_i
\end{equation}
with the $\{K_i\}$ so-called Kraus operators acting only on the $F-B$ sector, satisfying
\begin{equation}
\sum_i K_i K_i^\dag = \mathbb{I}_{FB}.
\end{equation}

Next, we ask: how are the reduced density matrices of Alice's particle on $\Sigma_1$ and $\Sigma_3$ related? To answer, we compute
\begin{eqnarray}
\rho_A (\Sigma_3) & = & Tr_{FB} \left[ \sum_i K_i^\dag \ket{\psi (\Sigma_1)}_{AFB} \bra{\psi (\Sigma_1)}_{AFB} K_i \right] \nonumber \\
& = & \sum_i Tr_{FB} \left[ K_i^\dag \ket{\psi (\Sigma_1)}_{AFB} \bra{\psi (\Sigma_1)}_{AFB} K_i \right] \nonumber \\
& = & \sum_i Tr_{FB} \left[ \ket{\psi (\Sigma_1)}_{AFB} \bra{\psi (\Sigma_1)}_{AFB} K_i K^\dag_i \right] \nonumber \\
& = & Tr_{FB} \left[ \ket{\psi (\Sigma_1)}_{AFB} \bra{\psi (\Sigma_1)}_{AFB} \sum_i K_i K^\dag_i \right] \nonumber \\
& = & Tr_{FB} \left[ \ket{\psi (\Sigma_1)}_{AFB} \bra{\psi (\Sigma_1)}_{AFB} \right] \nonumber \\
& = & \rho_A (\Sigma_1) , \nonumber 
\end{eqnarray}
where we used the cyclic property of the partial trace.

We see that the reduced density matrix of Alice's particle is exactly the same before and after the interaction between the field and any other objects. This, of course, means that interaction between the field and anything else is incapable of modifying any observations Alice could make on her particle, including the presence or absence of interference. We conclude that, if Alice performs her experiment in such a way that interference would be present in the absence of interaction between the field and other objects, the presence of any such objects would not modify the interference pattern in any way, fully closing the possibility of superluminal signaling in this type of experimental setups.

Besides relaxing the purely unitary evolution condition, there is another important sense in which our proof is more general than that in \cite{danielson}. In such an analysis, it is assumed that the initial state of the whole scenario, i.e., the state on $\Sigma_1$ of Alice, Bob and the field, is such that Bob's system is fully unentangled with the rest (see, e.g., Eq. (\ref{AB})). This however, might not be a good representation of the actual physical situation of the experiment. The point is that the act of Alice preparing her particle in the required superposition, has the potential of causing her field to get entangled, not only with Bob, his particle, and the trap, but also with a myriad of other objects in her vicinity. It is not obvious, then, that disregarding such a potential initial entanglement could not lead to problems with the analysis. Our proof, on the other hand, does not require such an assumption, as the initial state, $\ket{\psi (\Sigma_1)}_{AFB}$ is fully arbitrary. What we prove, then, is that, regardless of a potential initial entanglement between Bob, the field, Alice and anything else, further interaction between them does not alter in any way any observations made by Alice. This, however, does not mean that Alice would observe decoherence, regardless of the details of the initial state. Time evolution of the state does not affect Alice's observations, but different initial states could mean different levels of decoherence for Alice's interference experiment. 

For instance, different setups, involving a different number (and kind) of particles used by Bob, could correspond to different levels of decoherence affecting Alice's experiment. Of course, in any of these settings, the actions of Bob make no difference, but the initial setup very well might. All this is consistent with our proof because the transition between these different setups does not correspond to a physical process in time of the sort described by the analysis provided above.\footnote{What we mean here is that if one wanted to change the setup where one has, say, one particle and one trap, to one with $N$ particles and the corresponding traps, one would presumably have to consider bringing the new $ N-1$ particles in their traps from infinity to the neighborhood of the experimental region, and to consider the entanglement taking place between such devices and Alice's particle (and the one device already nearby).} In fact, the difference between setups evidently plays a role in accounting for the fact that some experimental settings work well in exhibiting interference, but others are ``spoiled'' by decoherence.

Before concluding, and given that the proof offered is somehow abstruse, we feel it convenient to include a few comments regarding the fundamental quantum features behind it. The basic quantum characteristic underlying the proof seems to be related to the way entanglement is distributed among a tripartite system. In particular, if X is entangled with Y, and Y interacts with Z (but not directly with X), then such an interaction might get Z entangled with X, but it cannot increase the entanglement that X is involved in. That is, entanglement initially shared between X and Y might now be distributed among entanglement between X and Y, and between X and Z, but the total amount of entanglement in which X in involved would not grow \cite{DE}. This basic fact, together with the spacetime distribution of the different elements of the experimental setting, ensure that impossibility of superluminal signaling.

Summing up, according to \cite{belenchia2018quantum}, the fields corresponding to the two paths of Alice's particle, initially, are nearly orthogonal, but if Alice recombines her particle slowly, the fields undergo ``false decoherence'', allowing for a perfect recombination. Above, we argued that this cannot be correct because, if the fields initially cause significant decoherence, then they would continue to do so on any spacial hypersurface containing the recombination of Alice's particle, fully destroying the interference pattern. Therefore, for Alice to observe interference, the fields must never cause significant decoherence. Still, it could seem that interaction between the fields and other objects could cause additional decoherence, opening a potential route for superluminal signaling. However, we have shown that interactions between the field and other objects cannot increase the amount of decoherence on Alice's particle. We conclude that, by analyzing the possible decoherence caused by the fields produced by Alice's particle and by paying attention to the way in which entanglement gets distributed in the gedankenexperiment, one can fully explain how superluminal signaling is averted.

\section{Conclusions}
\label{Co}

Recent works explore a gedankenexperiment in which the interaction of a particle with the field of a charged or massive object in a spatial quantum superposition, seems to allow for superluminal communication. Building on a previous analysis in \cite{mari2016experiments}, \cite{belenchia2018quantum} argues that, if one considers the quantization of the radiation of the fields in question, together with the presence of quantum vacuum fluctuations of such fields, then the alleged signaling disappears. Moreover, in the gravitational case, the result that quantization and vacuum fluctuations of the gravitational field are required to avoid signaling is promoted as an argument in favor of the necessity to quantize the gravitational field.

More recently, \cite{danielson} offers a more precise analysis, which is argued to improve on the rough estimates in \cite{belenchia2018quantum} and to show their conclusions to be valid in much more general scenarios.

In this work, we have identified a number of potential misunderstandings in some aspects of the aforementioned analyses and have provided what we take to be a complete account of the reasons behind the impossibility for superluminal signaling in these experiments. In particular, we have shown that attention to the way in which entanglement gets distributed in the experiment is enough to explain away the possibility of superluminal signaling.

\appendix
\section{Particles on a sphere}
\label{Ap}

In this appendix, we show that, if the analysis in \cite{belenchia2018quantum} were correct, then the decoherence caused by a set of particles arranged over a sphere, with a given density of particles per unit area, could be made independent of the radius of the sphere. To do so, consider a set of traps arranged over a sphere of radius $l$, centered on Alice. Assume all traps to contain particles of mass $m_B$ and charge $q_B$ and assume a uniform density $\rho$ of traps per unit area of the sphere. According to \cite{belenchia2018quantum}, the effect of Alice's field on the particles on the traps can be estimated by considering the ``effective dipole moment'' generated by Alice's superposition, which corresponds to an electric field with magnitude nowhere smaller than $q_A d/l^3$. Therefore, if liberated, the displacements of the particles on the traps would satisfy
\begin{equation}
\delta x \geq \frac{q_{B}}{m_{B}} \frac{\mathcal{D}_{A}}{l^{3}} T_{B}^{2} ,
\end{equation}
with $\mathcal{D}_{A} = q_A d$. Now, for a superluminal signal to be possible, $T_B$ would need to be smaller than $l$. Therefore, we take $T_B = \phi l$, with $\phi<1$, in which case
\begin{equation}
\delta x \geq \frac{q_{B}}{m_{B}} \frac{\mathcal{D}_{A}}{l} \phi^2 .
\end{equation}

Next, for simplicity, we assume the wave functions of the particles to be Gaussians of width $q_B/m_B$, which, according to \cite{belenchia2018quantum}, corresponds to the smallest possible size for such particles. In that case, the inner product of the left and right components of each particle would satisfy
\begin{equation}
\braket{L|R} \le e^{- \frac{\mathcal{D}_A^2 \phi^4}{8 l^2}}.
\end{equation}
Finally, taking into account that there are $N=4 \pi l^2 \rho$ particles on the sphere, the inner product of the left and right components of all of the particles would obey
\begin{equation}
\braket{L|R}^N \le e^{- \frac{1}{2}\pi \mathcal{D}_A^2 \phi^4 \rho},
\end{equation}
which is independent of $l$. Therefore, if $\rho$ and $\phi$ are kept constant, according to the analysis in \cite{belenchia2018quantum}, the decoherence caused by a given sphere would be independent of its radius. This means that, by employing enough such spheres, the decoherence caused on Alice's particle could be made as large as desired. 

Note that, since we took $T_B = \phi l$, farther particles would be allowed to react to Alice's field for a longer period of time. However, since we also took $\phi<1$, all such experiments would be spacelike separated from Alice. Note also that, if instead of covering the entire sphere, particles are arranged only over a patch of the sphere corresponding to a given solid angle $\Omega$, it would also be the case that, for constant $\rho$, $\phi$ and $\Omega$, the effect of those particles would be independent of the radius of the sphere. Therefore, a decoherence as large as desired could be caused, even if all of the traps were confined to a cone of arbitrarily small solid angle. Finally, note that the decision to release or not the particles from the traps could be coordinated by sending a light signal inwards from the outermost sphere, in such a way that even the last release is spacelike separated from Alice. Another option for coordination would be to distribute among all of the managers of the traps one of a set of spin-$\frac{1}{2}$ particles in the entangled state
\begin{equation}
\frac{1}{\sqrt{2}} \left( \ket{\uparrow}_1 \ket{\uparrow}_2...+\ket{\downarrow}_1 \ket{\downarrow}_2... \right)
\end{equation}
and to link the decision to release or not the particle of a given trap to the result of a spin measurement on the corresponding spin-$\frac{1}{2}$ particle, made by each manager at an appropriate time to ensure all experiments are spacelike separated from Alice.

\section*{Acknowledgments}

We would like to thank Víctor Torres Brauer and an anonymous referee for helpful comments. We also thank Daine L. Danielson, Gautam Satishchandran, Robert M. Wald for a valuable exchange. We acknowledge support from CONACYT grant 140630. DS acknowledges partial financial support from PAPIIT-DGAPA-UNAM project IG100120 and the grant FQXI-MGA-1920 from the Foundational Questions Institute and the Fetzer Franklin Fund, a donor advised by the Silicon Valley Community Foundation.


\bibliographystyle{ieeetr}

\bibliography{PrimerborradorRef.bib}


\end{document}